\documentclass[pra,floatfix]{revtex4}
\usepackage{amssymb}
\usepackage{amsmath}
\usepackage{graphicx}
\usepackage{subfigure}
\usepackage{float}
\usepackage{placeins}

\begin{document}

\title{Spontaneous symmetry breaking of photonic and matter waves in
two-dimensional pseudopotentials}
\author{Thawatchai Mayteevarunyoo$^{1}$, Boris A. Malomed$^{2,3}$ and
Athikom Reoksabutr$^{1}$}
\affiliation{$^{1}$Department of Telecommunication Engineering, Mahanakorn University of
Technology, Bangkok 10530, Thailand \\
$^{2}$Department of Physical Electronics, School of Electrical Engineering,
Faculty of Engineering, Tel-Aviv University, Tel-Aviv 69978, Israel\\
$^{3}$ICFO-Institut de Ciencies Fotoniques, Mediterranean Technology Park,
08860 Castelldefels (Barcelona), Spain\thanks{%
temporary Sabbatical address}}

\begin{abstract}
We introduce the two-dimensional Gross-Pitaevskii/nonlinear-Schr\"{o}dinger
(GP/NLS) equation with the self-focusing nonlinearity confined to two
identical circles, separated or overlapped. The model can be realized in
terms of Bose-Einstein condensates (BECs) and photonic-crystal fibers.
Following the recent analysis of the spontaneous symmetry breaking (SSB) of
localized modes trapped in 1D and 2D double-well nonlinear \ potentials
(also known as \textit{pseudopotentials}), we aim to find 2D solitons in the
two-circle setting, using numerical methods and the variational
approximation (VA). Well-separated circles support stable symmetric and
antisymmetric solitons. The decrease of separation $L$ between the circles
leads to destabilization of the solitons. The symmetric modes undergo two
SSB transitions. First, they are transformed into weakly asymmetric
breathers, which is followed by a transition to single-peak modes trapped in
one circle. The antisymmetric solitons perform a direct transition to the
single-peak mode. The symmetric solitons are described reasonably well by
the VA. For touching ($L=0$) and overlapping ($L<0$) circles, single-peak
solitons are found---asymmetric ones, trapped in either circle, and
symmetric solitons centered at the midpoint of the bi-circle configuration.
If the overlap is weak, the symmetric soliton is unstable. It may
spontaneously leap into either circle and perform shuttle motion in it. A
region of stability of the symmetric solitons appears with the increase of
the overlap degree. In the case of a moderately strong overlap, another SSB
effect is found, in the form of a pair of symmetry-breaking and restoring
bifurcations which link families of the symmetric and asymmetric solitons.

\end{abstract}

\maketitle

\section{Introduction}

A large number of systems in optics and atomic physics feature a dual-core
(double-well) structure, based on an effective potential in the form of a
symmetric set of two minima (wells). The wells are linked by the tunneling
of optical waves or matter waves, trapped in them, across the barrier
separating the wells. In cases when the trapped waves are subject to
nonlinear self-interaction, a fundamental property of the symmetric
double-well systems, which originates from the competition between the
linear tunnel coupling between the wells and the nonlinearity acting inside
of them, is the \textit{spontaneous symmetry breaking} (SSB). In particular,
the SSB of solitons (solitary waves trapped in the double-well potential)
was first studied in the model of dual-core optical fibers \cite{Skinner}-%
\cite{Progress}. A similar analysis was carried out for Bose-Einstein
condensates (BECs) loaded into double-well potentials \cite{mean-field}-\cite%
{mean-field5}. In the experiment, the SSB has been observed in the BEC with
repulsive interactions between atoms \cite{Heidelberg}, and in an optical
setting based on photorefractive crystals \cite{ZChen}. It has been
demonstrated that, in the systems with attractive or repulsive intrinsic
nonlinearities, stationary asymmetric modes are generated by \textit{%
symmetry-breaking} \textit{bifurcations} from symmetric or antisymmetric
trapped solitons, respectively. SSB effects for solitons were also studied
in the model with competing nonlinearities of both these types, \textit{viz}%
., self-focusing cubic and self-defocusing quintic nonlinear terms, which
gives rise to closed \textit{bifurcation loops} \cite{CQ,CQ3}.

The analysis of the SSB in matter-wave (BEC) settings and their photonic
counterparts was extended for two-dimensional (2D) solitons, supported by
the interplay of the self-attractive \cite{Warsaw}-\cite{Arik3} or
self-repulsive \cite{Arik}-\cite{Markus} nonlinearity and spatially periodic
potentials acting in each core. The periodic potential is necessary to
stabilize 2D solitons (in each core separately) against the collapse in the
case of the self-attraction \cite{review}, or support solitons of the gap
type in the case of the self-repulsion \cite{gap1}-\cite{gap3}. The analysis
of the SSB in a 2D localized configuration of a different type, based on the
set of four potential wells forming a square pattern (in a single core) was
reported in Ref. \cite{square}.

Photonic and matter waves can be trapped not only in the usual
linear potentials, but also in nonlinear \textit{pseudopotentials}
(this term is often used in solid-state physics \cite{Harrison}),
which may be induced by a spatial modulation of the local
nonlinearity coefficient. In BEC, the \ local nonlinearity may be
controlled, through the Feshbach-resonance effect, by an external
magnetic field \cite{Feshbach}. Accordingly, the spatial
nonlinearity modulation may be induced by an inhomogeneous
magnetic field \cite{SU}. In optics, similar settings can be
designed as all-solid or liquid-filled microstructured fibers,
using combinations of materials with matched refractive indices
but different Kerr coefficients, as proposed in Ref.
\cite{Barcelona-OL}.

The studies of soliton dynamics in purely nonlinear (pseudo)potentials, as
well as in their combinations with usual linear potentials, has recently
grown into a vast research area, see the review article \cite{Barcelona-RMP}%
. Models with various patterns of the spatial nonlinearity modulation were
studied in detail in 1D geometries \cite{NL}-\cite{Barcelona-vector}, \cite%
{Barcelona-RMP}. In 2D, the analysis is essentially more challenging, as, in
this case, it is difficult to stabilize localized modes against the
spatiotemporal \textit{collapse }(catastrophic self-focusing induced by the
cubic nonlinearity),\textit{\ }using only the local modulation of the
nonlinearity, without the support provided by a linear potential \cite{HS}-%
\cite{Hung}, \cite{Barcelona-OL}, \cite{Barcelona-RMP}. Nevertheless, 2D
nonlinear structures capable to support stable solitons have been found.
Basic types of such structures are represented by a circle or annulus,
filled with the self-focusing material, which is embedded into a linear or
defocusing medium \cite{HS} (as well as a lattice of such circles \cite%
{Barcelona-OL}), and a single or double stripe made of the same material
\cite{Hung}. In both cases, it was concluded that the
nonlinearity-modulation profiles with sharp edges provide for much more
efficient stabilization of 2D solitons than similar smooth profiles. On the
other hand, no pattern of 2D modulations of the self-focusing cubic
nonlinearity was found that would support stable vortex solitons.

Coming back to 1D, it was found that the simplest pseudopotential capable to
sustain stable localized modes is built as a symmetric set of two
delta-functions multiplying the cubic self-focusing nonlinearity \cite{we}.
Explicit analytical solutions are available for all stationary modes trapped
in this setting -- symmetric, antisymmetric, and asymmetric ones. The
analytical solution allows one to study the corresponding SSB bifurcation,
which generates asymmetric modes from the symmetric ones. On the other hand,
the exact solution of the 1D model with two delta-functions is partly
degenerate, which, in particular, makes the asymmetric modes completely
unstable. The replacement of the ideal delta-functions by their regularized
counterparts lifts the degeneracy, making the asymmetric modes partly or
fully stable \cite{we}.

The model with a pair of ideal or regularized delta-functions may be used as
a paradigm for the study of the SSB of localized modes supported by
symmetric double-well pseudopotential structures. The objective of the
present work is to introduce a natural generalization of this setting in the
2D geometry, with the nonlinearity concentrated in two separated or
overlapping identical circles with sharp edges, see Fig. \ref{fig1} and
insets to Figs. \ref{fig9}(a) and \ref{fig13}(a) below (ideal $\delta $%
-functions are irrelevant in the 2D case, see Eqs. (\ref{Poisson}), (\ref{ln}%
) and the related text in the next section). This setting can be realized in
BEC and nonlinear optics alike, by means of the same techniques which were
outlined above, i.e., respectively, the use of the Feshbach resonance,
controlled by a spatially nonuniform magnetic field, or a crystalline
structure formed by a pair if intermingled materials with equal refractive
indices but different Kerr coefficients.

The paper is organized as follows. The model is introduced in detail in
Section 2. Then, in Section 3 we develop the variational approximation (VA),
with the objective to describe patterns with different symmetries supported
by the symmetric pair of well-separated circles. Results of the numerical
analysis are collected in Section 4. The results describe the transition
from double-peak symmetric and antisymmetric modes to asymmetric ones, with
the decrease of the separation between the circles, which implies the
strengthening interaction between them. The symmetric solitons perform the
symmetry-breaking transition via an intermediate mode in the form of weakly
asymmetric breathers, while antisymmetric solitons directly jump into
strongly asymmetric single-peak modes. For touching or partly overlapping
circles, only single-peak solitons exist. These may be either symmetric
ones, centered at the midpoint of the bi-circle configuration, or solitons
spontaneously leaping into either circle and then featuring shuttle motion
in that trap. Various manifestations of the SSB in the present system are
summarized in Section 4, which concludes the paper.

\section{The model}

\subsection{The equations}

\begin{figure}[tbph]
\centering\includegraphics[width=3in]{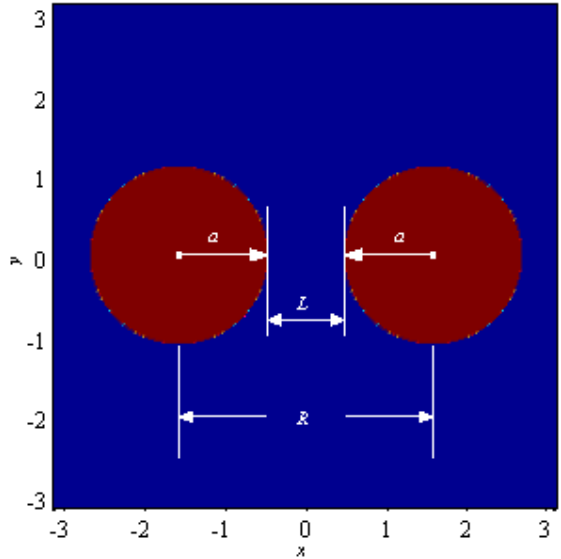}
\caption{(Color online) The geometry of the bi-circle system.}
\label{fig1}
\end{figure}

The setting considered in this work is shown in Fig. \ref{fig1}, where $a$
is the radius of the two identical circles, and $L$ is the separation
between them, $R\equiv L+2a$ being the distance between the centers of the
circles, which are located at points $\left\{ x_{0},y_{0}\right\} =\left\{
\pm \left( L/2+a\right) ,0\right\} $. Along with the configuration displayed
in Fig. \ref{fig1}, we will also consider the one with $R<2a$ (i.e., $L<0$),
which corresponds to partly overlapping circles [see top left insets to
Figs. \ref{fig9}(a) and \ref{fig13}(s) below].

The scaled form of the underlying Gross-Pitaevskii/nonlinear-Schr\"{o}dinger
(GP/NLS) equation with the self-focusing ($g_{1}>0$) nonlinearity confined
to the two circles is%
\begin{equation}
i\psi _{t}=-\left( \psi _{xx}+\psi _{yy}\right) -g\left( x,y\right)
\left\vert \psi \right\vert ^{2}\psi ,  \label{psi}
\end{equation}%
where the simplest form of the localization may be chosen as%
\begin{equation}
g(x,y)=g_{1}\left[ \exp \left( -\frac{\left( x-R/2\right) ^{2}+y^{2}}{a^{2}}%
\right) +\exp \left( -\frac{\left( x+R/2\right) ^{2}+y^{2}}{a^{2}}\right) %
\right] .  \label{g}
\end{equation}%
In the case of the GP equation, $t$ is normalized time, while in optics, $t$
is actually the propagation distance in the bulk waveguide. In either case, $%
x$ and $y$ are the transverse coordinates. In fact, a different form of the
2D modulation function will be used for numerical calculations, see Eqs. (%
\ref{g1(x,y)}-(\ref{g(x,y)} below.

Stationary solutions to Eq. (\ref{psi}), with chemical potential $\mu $ of
the BEC (or propagation constant $-\mu $ in the optical waveguide) are
sought for as $\psi \left( x,y\right) =e^{-i\mu t}\phi \left( x,y\right) $,
where real function $\phi \left( x,y\right) $ obeys equation%
\begin{equation}
\mu \phi +\left( \phi _{xx}+\phi _{yy}\right) +g\left( x,y\right) \phi
^{3}=0.  \label{phi}
\end{equation}%
Equation (\ref{phi}) will be solved below by means of a numerical method,
and the VA will be used too.

\subsection{The Lagrangian structure and variational ansatz}

Equation (\ref{phi}) can be derived from the Lagrangian, $L=\int \int
\mathcal{L}dxdy$, with density%
\begin{equation}
2\mathcal{L}=-\mu \phi ^{2}+\left( \phi _{x}\right) ^{2}+\left( \phi
_{y}\right) ^{2}-(1/2)g\left( x,y\right) \phi ^{4}.  \label{density}
\end{equation}%
Obviously, localized solutions have $\mu <0$. The energy corresponding to
Eq. (\ref{psi}) is%
\begin{equation}
E=\int \int \left[ \left( \left\vert \psi _{x}\right\vert ^{2}+\left\vert
\psi _{y}\right\vert ^{2}\right) -\frac{1}{2}g\left( x,y\right) \left\vert
\psi \left( x,y\right) \right\vert ^{4}\right] dxdy.  \label{E}
\end{equation}

The application of the VA is possible in the case when $a$ in Eq. (\ref{g})
is small, hence $g\left( x,y\right) $ may be approximated \ (only for the
purpose of the development of the VA) by a combination of delta-functions:%
\begin{equation}
g\left( x,y\right) =G\left[ \delta \left( x-R/2\right) \delta \left(
y\right) +\delta \left( x+R/2\right) \delta (y)\right] ,~G\equiv \pi
g_{1}a^{2},  \label{delta}
\end{equation}%
where coefficient $G$ is determined by the integral-balance condition, $\int
\int g\left( x,y\right) dxdy\equiv G\int \int \left[ \delta \left(
x-R/2\right) \delta \left( y\right) +\delta \left( x+R/2\right) \delta (y)%
\right] dxdy$. The VA developed below is based on the use of the following
\textit{ansatz},%
\begin{equation}
\phi \left( x,y\right) =\exp \left( -\frac{2y^{2}}{R^{2}l^{2}}\right)
\left\{ A\exp \left[ -\frac{2\left( x-R/2\right) ^{2}}{R^{2}l^{2}}\right]
+B\exp \left[ -\frac{2\left( x+R/2\right) ^{2}}{R^{2}l^{2}}\right] \right\} ,
\label{ansatz}
\end{equation}%
where $A,B$ and $l$ are variational parameters, the latter one being the
width of the ansatz measured in units of $R/2$. The ansatz is isotropic in
the vicinity of each attractive center (the variational equations for an
anisotropic ansatz turn out to be very cumbersome). It is assumed that the
asymmetry between the wave functions in vicinities of the two attractive
centers may be accounted for by a difference in the amplitudes, $A\neq B$,
while widths $m$ and $l$ are assumed to be the same near both centers.

Comparing expression (\ref{ansatz}) with Eqs. (\ref{g}) demonstrates that
approximation (\ref{delta}), based on the delta-functions, may be valid for
sufficiently small $a$ in Eq. (\ref{g})---namely, if the VA will eventually
yield
\begin{equation}
l\gg 2a/R.  \label{a}
\end{equation}%
On the other hand, it is easy to check that Eq. (\ref{phi}) with the
nonlinearity-modulation function represented by the single 2D
delta-function, i.e., $g\left( x,y\right) =G\delta (x)\delta (y)$, does not
give rise to any soliton solution (unlike its 1D counterpart \cite{we}).
Indeed, taking the corresponding equation,%
\begin{equation}
\mu \phi +\left( \phi _{xx}+\phi _{yy}\right) +G\delta \left( x)\delta
(y\right) \phi ^{3}=0,  \label{phi-delta}
\end{equation}%
and setting, at first, $\mu =0$, one can use the commonly known fact that
the Poisson equation in the form of
\begin{equation}
\phi _{xx}+\phi _{yy}=C\delta (x)\delta (y),  \label{Poisson}
\end{equation}%
with constant $C$, has the fundamental solution,
\begin{equation}
\phi (r)=\left( C/2\pi \right) \ln \left( r/r_{0}\right) ,  \label{ln}
\end{equation}%
where $r\equiv \sqrt{x^{2}+y^{2}}$, and $r_{0}$ is an arbitrary positive
constant. Temporarily introducing a small cutoff radius, $\rho $, formula (%
\ref{ln}) suggests that a solution to Eq. (\ref{phi-delta}) can be sought
for as $\phi (r)=\phi _{0}\ln \left( r/r_{0}\right) $ , which corresponds to
$C=-G\left( \phi \left( r=\rho \right) \right) ^{3}=G\phi _{0}^{3}\left( \ln
\left( r/\rho _{0}\right) \right) ^{3}\equiv C_{\rho }$ in Eq. (\ref{Poisson}%
). Then, the self-consistency condition ensuing from Eq. (\ref{ln}), $\phi
_{0}=C_{\rho }/2\pi ,$ yields $\phi _{0}^{2}=\left( 2\pi /G\right) \left[
\ln \left( r_{0}/\rho \right) \right] ^{-3}.$This relation shows that the
limit of $\rho \rightarrow 0$ corresponds to $\phi _{0}\rightarrow 0$, which
means that solely the trivial solution, $\phi \equiv 0$, is possible if the
cutoff is removed.

A similar conclusion [the nonexistence of nontrivial solutions to Eq. (\ref%
{phi-delta})] can be obtained for $\mu <0$, using an appropriate Hankel's
function, instead of solution (\ref{ln}). Thus, the use of approximation (%
\ref{delta}) in the framework of the VA is meaningful for the circles of a
small but finite radius $a$, which obeys condition (\ref{a}).

\section{The variational analysis}

The substitution of ansatz (\ref{ansatz}) into density (\ref{density}) and
calculation of the integrals yields the following expression for the
effective Lagrangian:%
\begin{gather}
L_{\mathrm{eff}}=-\left( \pi /8\right) R^{2}\mu l^{2}\left(
A^{2}+B^{2}+2e^{-1/l^{2}}AB\right)  \notag \\
+\left( \pi /2\right) \left[ \left( A^{2}+B^{2}\right) +2\left(
1-l^{-2}\right) e^{-1/l^{2}}AB\right]  \notag \\
-G\left( R/4\right) ^{2}\left[ \left( 1+e^{-8/l^{2}}\right) \left(
A^{4}+B^{4}\right) \right.  \notag \\
\left. +4e^{-2/l^{2}}\left( 1+e^{-4/l^{2}}\right) AB\left(
A^{2}+B^{2}\right) +12e^{-4/l^{2}}A^{2}B^{2}\right] ,  \label{simple}
\end{gather}%
the norm of the ansatz being%
\begin{equation}
N\equiv \int \int \phi ^{2}(x,y)dxdy=\pi \left( Rl/2\right) ^{2}\left(
A^{2}+B^{2}+2e^{-1/l^{2}}AB\right) .  \label{N}
\end{equation}%
In particular, for the symmetric and antisymmetric solitons, with $A=\pm B$,
Eqs. (\ref{simple}) and (\ref{N}) give%
\begin{gather}
L_{\mathrm{eff}}^{\left( \pm \right) }=\pi A^{2}\left[ -\left( R/2\right)
^{2}\mu l^{2}\left( 1\pm e^{-1/l^{2}}\right) +1\pm \left( 1-l^{-2}\right)
e^{-1/l^{2}}\right]  \notag \\
-\left( G/8\right) R^{2}A^{4}\left[ 1+e^{-8/l^{2}}\pm 4e^{-2/l^{2}}\left(
1+e^{-4/l^{2}}\right) +6e^{-4/l^{2}}\right] ,  \label{+-}
\end{gather}%
\begin{equation}
N=\pi l^{2}\left( R^{2}/2\right) \left( 1+e^{-1/l^{2}}\right) A^{2}~.
\label{N-symm}
\end{equation}

Even after the simplifications, the Euler-Lagrange equations, $\partial L_{%
\mathrm{eff}}^{\left( \pm \right) }/\partial \left( A^{2}\right) =\partial
L_{\mathrm{eff}}^{\left( \pm \right) }/\partial \left( l^{2}\right) =0$,
following from Eq. (\ref{+-}) remain cumbersome. A tractable approximation
can be developed if $l^{2}$ is small enough to neglect exponentially small
factors $e^{-4/l^{2}}$ and $e^{-8/l^{2}}$ in comparison with $1$ in
expression (\ref{+-}) (in fact, we will then have $l^{2}<1/2$, see below).
In this approximation, the variational equations for the symmetric solitons
are%
\begin{equation}
\mu \left[ 1+\left( l^{-2}+1\right) e^{-1/l^{2}}\right] =-\left( 2/R\right)
^{2}l^{-4}\left( l^{-2}-2\right) e^{-1/l^{2}},  \label{mu-symm}
\end{equation}%
\begin{equation}
A^{2}=\frac{\pi \left[ 1-\left( l^{-2}-1\right) e^{-1/l^{2}}-(R/2)^{2}\mu
l^{2}\left( 1+e^{-1/l^{2}}\right) \right] }{G\left( R/2\right) ^{2}\left(
1+4e^{-2/l^{2}}\right) }.  \label{A-symm}
\end{equation}%
It follows from Eq. (\ref{mu-symm}) that the localization condition, $\mu <0$%
, entails $l^{2}<1/2$, as mentioned above. Then, both terms in the square
brackets in Eq. (\ref{A-symm}) are positive, hence $A^{2}$ is positive too,
as it should be.

An explicit analysis is also possible for the limit case of very broad
solitons, $l^{2}\gg 1$. In this case, the expansion of Lagrangian (\ref%
{simple}) in powers of $l^{-2}$ yields%
\begin{eqnarray}
L_{\mathrm{eff}} &\approx &-\left( \pi /8\right) R^{2}\mu l^{2}\left(
A+B\right) ^{2}  \notag \\
&&+\left( \pi /2\right) \left( A+B\right) ^{2}-G\left( R^{2}/8\right) \left(
A+B\right) ^{4}  \notag \\
&&+l^{-2}\left[ -2\pi AB+G\left( R^{2}/2\right) \left( A+B\right) ^{4}\right]
.  \label{simplest}
\end{eqnarray}%
A straightforward consideration of the the Euler-Lagrange equations
following from Eq. (\ref{simplest}) demonstrates that they give rise to no
antisymmetric solutions, while the symmetric solution is found in the
following form, valid at the lowest order in $l^{-2}$:%
\begin{eqnarray}
A^{2} &=&\ B^{2}=\pi /\left( 2GR^{2}\right) ,~N=\left[ \pi /\left( 2G\right) %
\right] ^{2}l^{2},  \label{const} \\
N &=&\left( \pi ^{2}/GR\right) \left( -\mu \right) ^{-1/2}.  \label{asympt}
\end{eqnarray}%
In this approximation, Eq. (\ref{const}) demonstrates that the amplitude of
the symmetric soliton is constant, while its norm diverges $\sim l^{2}$, and
the entire soliton family is described by dependence $N(\mu )$ given by Eq. (%
\ref{asympt}).

\section{Numerical results}

\subsection{The setting}

For the numerical analysis, we return from approximation (\ref{delta}) to
the configuration displayed in Fig. \ref{fig1}, and define the nonlinearity
as follows [cf. Eq. (\ref{g})]:
\begin{equation}
g_{1}\left( x,y\right) =\exp \left[ -\left( r_{L}/a\right) ^{k}\right] +\exp %
\left[ -\left( r_{R}/a\right) ^{k}\right] ,  \label{g1(x,y)}
\end{equation}%
\begin{equation}
g_{2}\left( x,y\right) =\left\vert \exp \left[ -\left( r_{L}/a\right) ^{k}%
\right] -\exp \left[ -\left( r_{R}/a\right) ^{k}\right] \right\vert
\label{g2(x,y)}
\end{equation}%
\begin{equation}
g(x,y)=(1/2)\left[ g_{1}\left( x,y\right) +g_{2}\left( x,y\right) \right] ,
\label{g(x,y)}
\end{equation}%
where $r_{L}\equiv \sqrt{y^{2}+\left( x+R/2\right) ^{2}}$ and $r_{R}\equiv
\sqrt{y^{2}+\left( x-R/2\right) ^{2}}$. As mentioned above, the
stabilization of 2D solitons supported by the single nonlinear circle is
facilitated by making edges of the circles sharp \cite%
{HS,Barcelona-OL,Barcelona-RMP}. To implement this feature in the present
setting, $k=100$ was used in Eqs. (\ref{g1(x,y)}) and (\ref{g2(x,y)}) [the
full nonlinearity (\ref{g(x,y)}) is then almost identical to $g_{1}\left(
x,y\right) $].

Localized stationary modes were found as solutions to Eq. (\ref{phi}) by
means of the so-called Newton-conjugate-gradient method \cite{method} (the
ordinary version of the Newton's method did not lead to convergent solutions
in the present setting). After finding the stationary solutions, their
stability was tested by means of direct simulations of Eq. (\ref{psi}),
performed by means of the usual split-step method.

In addition to presenting the numerical results, we also include some
predictions produced by the VA [chiefly, in Fig. \ref{fig2}(a)]. To this
end, norm\textbf{\ }(\ref{N-symm})\textbf{\ }was plotted vs.\textbf{\ }$\mu $%
, taking $A^{2}$\textbf{\ }as per Eq.\textbf{\ (}\ref{A-symm}\textbf{) }and
eliminating $l$, in favor of $\mu $, with the help of Eq. (\ref{mu-symm}).

\subsection{Separated circles ($L>0$)}

\subsubsection{Symmetric solitons}

\begin{figure}[tbph]
\centering\subfigure[]{\includegraphics[width=3.5in]{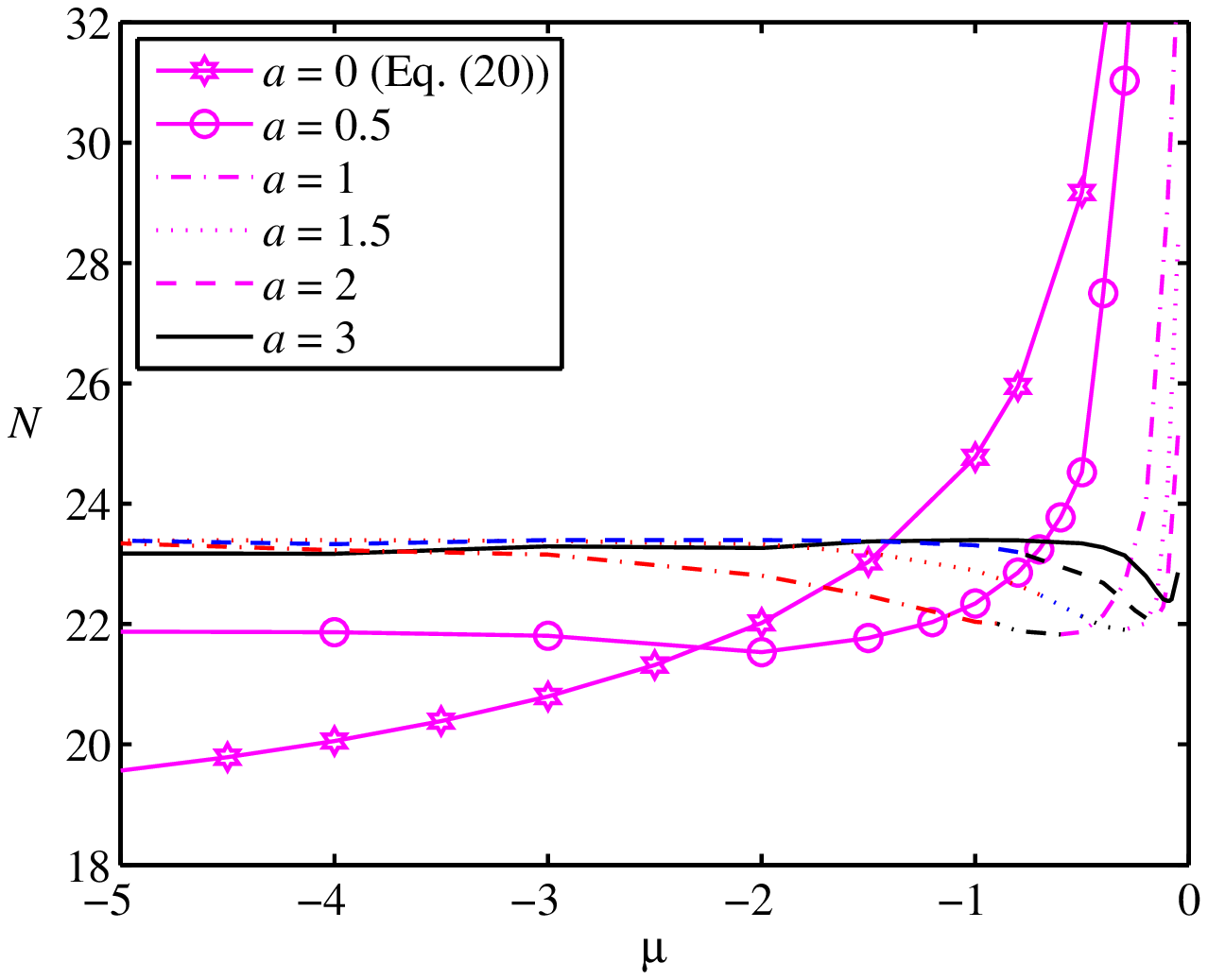}}%
\subfigure[]{\includegraphics[width=3.5in]{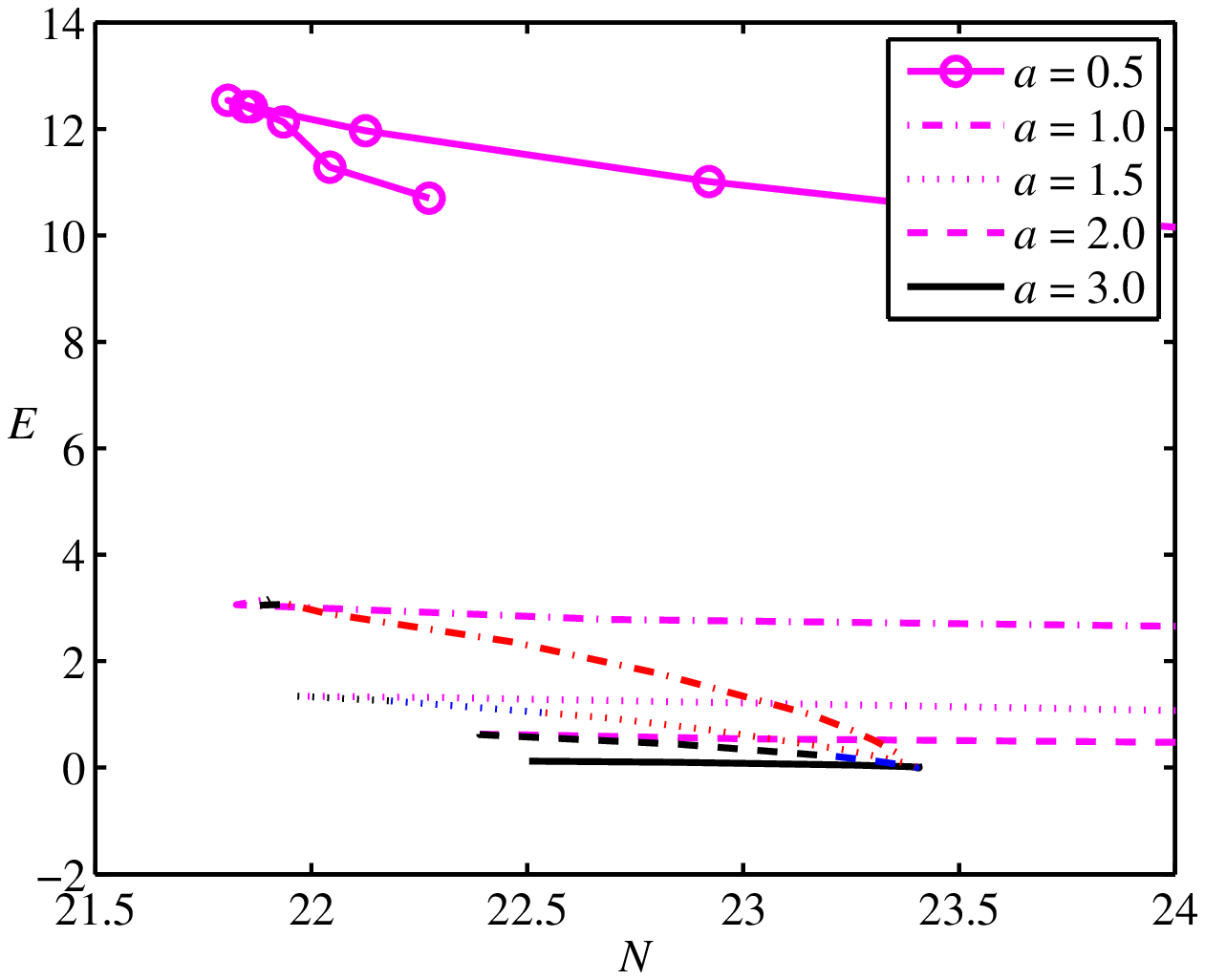}}
\caption{(Color online) Norm $N$ versus $\protect\mu $ (a) and energy $E$
versus $N$ (b) for symmetric solitons at different values of radius $a$ of
the circles, with a fixed separation between them, $L=3\protect\pi $. The
stability of the solution families is identified as follows: the magenta
color pertains to the instability through decay; black -- the instability
via a jump into a strongly asymmetric mode; red -- stability (it is a part
of the curve for $a=1$ with the negative slope); blue -- spontaneous
transformation of the stationary solitons into robust breathers.}
\label{fig2}
\end{figure}

We start by presenting results for symmetric modes, with $\phi \left(
x,y\right) =\phi \left( -x,y\right) $, in the case of $L>0$, i.e.,
non-overlapping circles, see Fig. \ref{fig1}. Figures \ref{fig2}(a) and (b)
display the overall characteristics of families of the symmetric modes, in
the form of dependences $N\left( \mu \right) $\ and $E(N)$, i.e., the norm
versus the chemical potential and the energy versus the norm [see Eqs. (\ref%
{N}) and (\ref{E})], for different values of radius $a$ of the circles and a
fixed separation between them, $L=3\pi $ (this value was selected to display
generic results).

The curve marked ``$a=0$" in Fig. \ref{fig2}(a) displays the prediction of
the VA for the symmetric solitons, obtained from Eqs. (\ref{N-symm})\textbf{%
, (}\ref{A-symm}\textbf{), }and (\ref{mu-symm}), as outlined above [recall
that the VA was developed for the nonlinearity modulation taken in the form
of Eq. (\ref{delta}), which formally corresponds to $a=0$]. The shape of the
VA-predicted curve is reasonably close to its numerical counterparts
generated for finite $a$. The other tractable approximation, which amounts
to Eq. (\ref{asympt}) for $\mu \rightarrow -0$ and $N\rightarrow \infty $,
correctly predicts the asymptotic form of the numerically found curves in
this limit.

The opposite limit of $\mu \rightarrow -\infty $ corresponds to very narrow
peaks pinned to the center of each circle. When the radius of these peaks is
much smaller than the radius of the circles ($a$), they are nearly identical
to the Townes solitons \cite{Berge}, whose norm (i.e., the collapse
threshold in the uniform medium with the self-focusing cubic nonlinearity)
is
\begin{equation}
N_{\mathrm{thr}}\approx 11.69.  \label{Townes}
\end{equation}%
This fact explains the limit value $N(\mu \rightarrow -\infty )\approx
23.5~\simeq 2N_{\mathrm{thr}}$ observed in Fig. \ref{fig2}(a) for $%
a=1,~1.5,~2,$ and $3$. The form of the VA adopted above is irrelevant in
this case, as approximation (\ref{delta}) implies that the radius of the
peak is much larger than $a$. However, the isolated peak in the medium with
the uniform cubic nonlinearity (i.e., the unstable Townes' soliton) may be
approximated by another version of the VA, which yields a reasonable
approximation for the collapse threshold, $N_{\mathrm{thr}}^{\left( \mathrm{%
VA}\right) }=4\pi $ \cite{Anderson}.

The plots displayed in Figs. \ref{fig2}(a) and \ref{fig2}(b) identify the
stability of the families of symmetric solitons. A well-known necessary, but
not sufficient, condition for the stability of solitons supported by the
self-focusing nonlinearity amounts to the Vakhitov-Kolokolov (VK) criterion,
$dN/d\mu <0$ \cite{Berge}. It is seen in Fig. \ref{fig2}(a) that this
criterion is indeed necessary but not sufficient, as the portion of the $%
N(\mu )$ curve with the negative slope is stable only for $a=1$. It is
worthy to note that, as seen in Fig. \ref{fig2}, the stable portion of the
latter curve corresponds to lower values of the energy than the unstable
part of the same curve.

\begin{figure}[tbph]
\centering\includegraphics[width=4in]{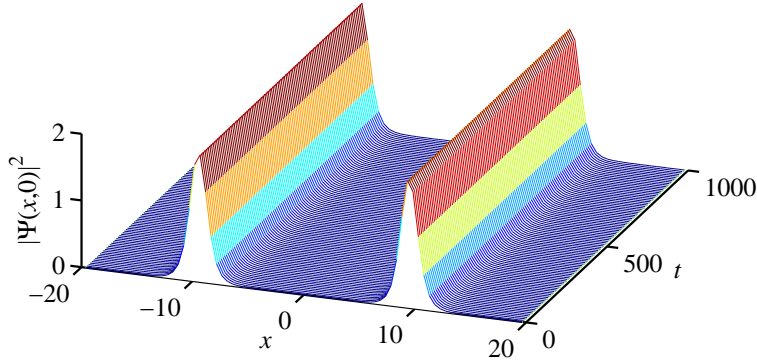}
\caption{(Color online) An example of a stable symmetric soliton, with norm $%
N=22$ ($\protect\mu =-0.23045$), $R=6\protect\pi $, and $a=1.5$. The
evolution of the soliton is shown in the $x$-cross section drawn through $%
y=0 $.}
\label{fig3}
\end{figure}

\begin{figure}[tbp]
\centering\subfigure[]{\includegraphics[width=4in]{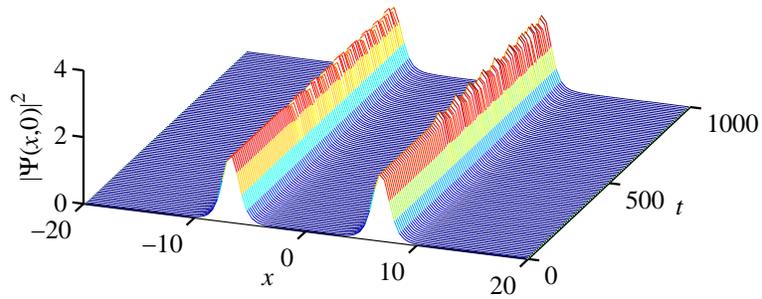}} %
\subfigure[]{\includegraphics[width=4in]{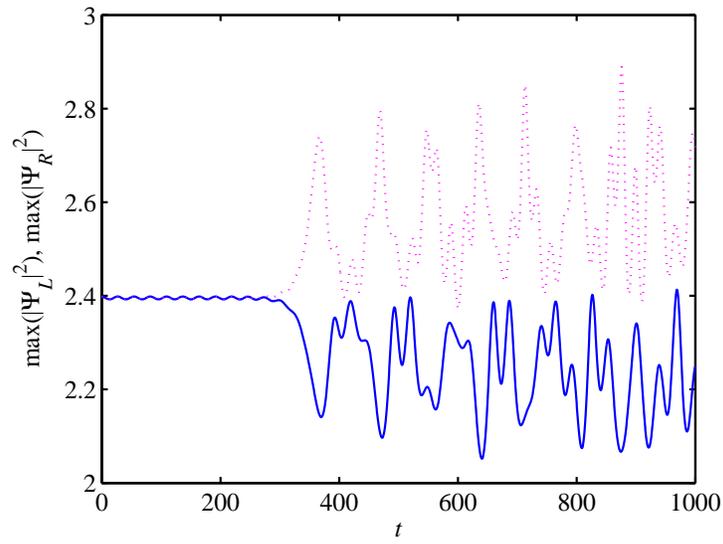}}
\caption{(Color online) (a) An example of the transformation of a weakly
unstable symmetric soliton into a robust breather, with $N=22$, $R=4.3%
\protect\pi $ and $a=1.5$. (b) The evolution of largest values of the local
density of the left (``L") and right (``R") peaks, which features a weak
spontaneous breaking of the symmetry between the peaks.}
\label{fig4}
\end{figure}

A typical example of a stable symmetric (\textit{double-peak}) soliton is
displayed in Fig. \ref{fig3} (it pertains to the separation between the
circles $L\equiv R-2a=6\pi -3\approx \allowbreak 15.85$, which is different
from $L=3\pi $ fixed in Fig. \ref{fig2}, but there is no essential
difference in the shape of the stable solitons in these cases). The
evolution of weakly unstable symmetric solitons leads to the formation of
breathers, as shown in Fig. \ref{fig4}. In this case, the soliton is
spontaneously transformed into a robust localized object featuring irregular
small-amplitude intrinsic oscillations. It is worthy to note that the
transition from the stationary solitons to breathers leads to a relatively
weak but tangible SSB effect, as seen in Fig. \ref{fig4}(b). Numerical data
demonstrate that the breather keeps practically the entire initial norm,
i.e., radiation losses are negligible in the course of the rearrangement of
the symmetric soliton into the breather.

\begin{figure}[tbp]
\centering\includegraphics[width=4in]{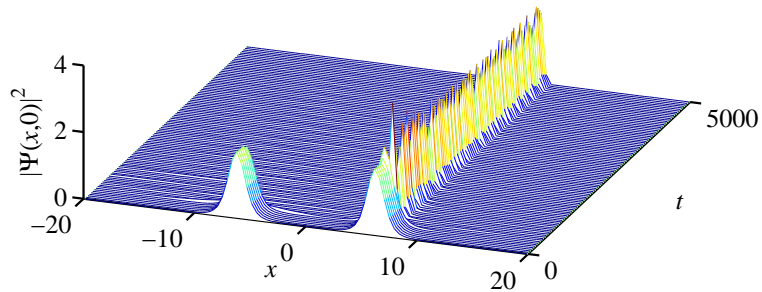}
\caption{(Color online) The spontaneous transformation of an unstable
symmetric double-peak soliton into a single-peak one, at $N=22$, $R=4\protect%
\pi $ and $a=1.2$.}
\label{fig5}
\end{figure}

The instability mode designated as the decay into radiation in Fig. \ref%
{fig2} means that the soliton completely disintegrates (not shown here in
detail). On the other hand, the evolution outcome marked as the instability
via a jump into a strongly asymmetric mode implies that the symmetric
double-peak soliton suffers strong SSB, which leads to a sudden transition
into a robust \textit{single-peak} mode nested in either circle, as shown in
Fig. \ref{fig5}. Note that, in the case displayed in Fig. \ref{fig2}, the
initial norm ($N=22$) is much higher than critical value (\ref{Townes}). The
numerical data demonstrate that the eventual value of the norm of the
emerging single-peak mode is slightly smaller than the threshold value (\ref%
{Townes}), the surplus norm, $\Delta N\simeq 10$, being shed off in the form
of radiation.

\begin{figure}[tbph]
\centering\includegraphics[width=4in]{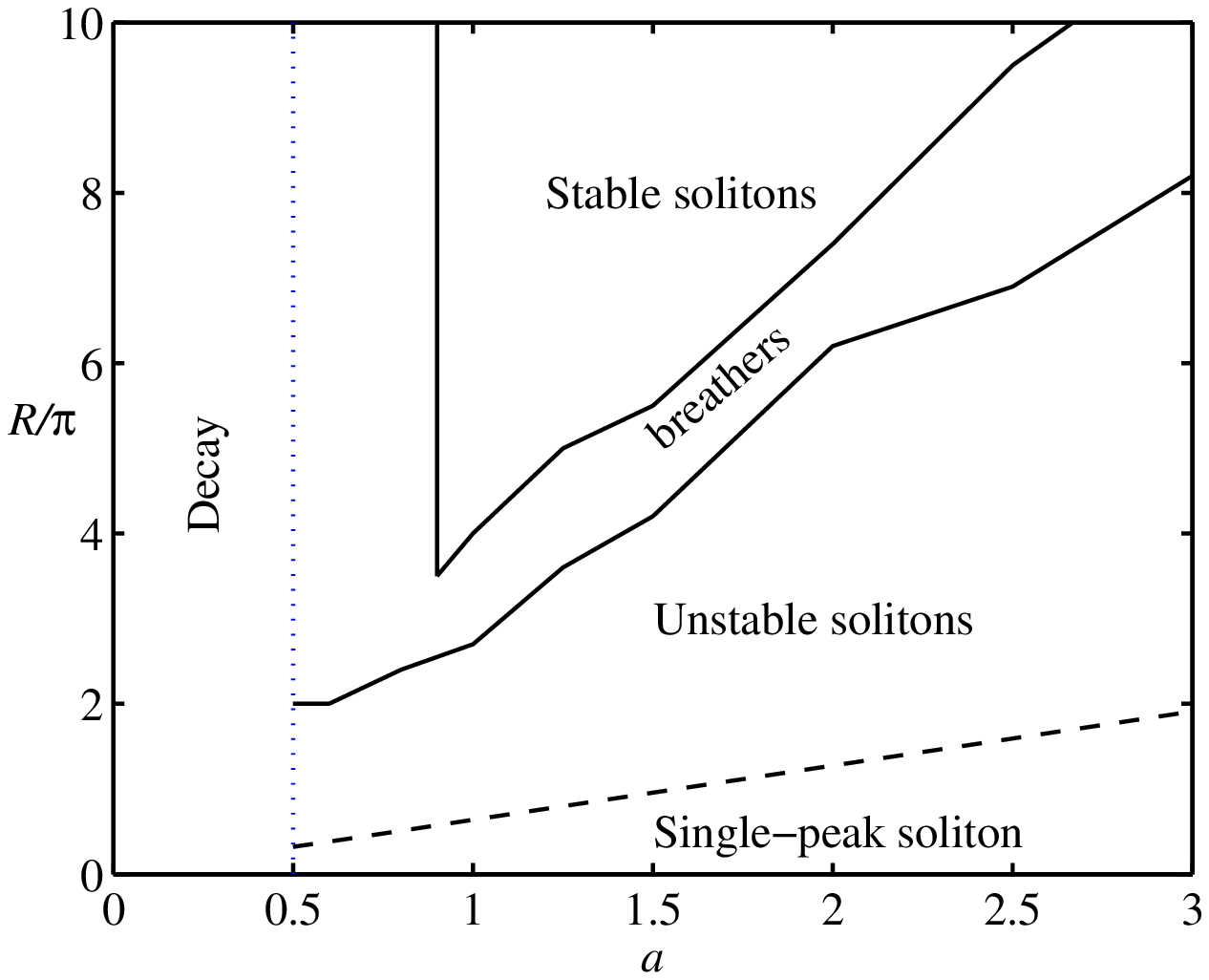}
\caption{(Color online) The stability diagram for various localized modes
developing from stationary symmetric solitons at a fixed value of the norm, $%
N=22$, in the plane of the radius of the circles ($a$) and normalized
separation between their centers ($R/\protect\pi $). ``Unstable solitons"
are those which spontaneously transform into the single-peak modes, as shown
in Fig. \protect\ref{fig5}. The dashed line represents $L=0$, i.e., $R=2a$.
Along this line, the two circles are tangent to each other, and below it the
circles partly overlap. As shown in the next section, only single-peak
solitons may be stable at $L\leq 0$.}
\label{fig6}
\end{figure}

The results of the numerical analysis of the stability of the symmetric
solitons is summarized in the diagram in the plane of $\left( a,R/\pi
\right) $, which is displayed in Fig. \ref{fig6}, for a fixed total norm, $%
N=22$. An obvious feature is that radius $a$ of the circles supporting
stable symmetric solitons must not be too small (roughly speaking, all the
solitons are unstable at $a<1$). In fact, stable symmetric solitons (as well
as antisymmetric ones, dealt with in the next subsection, see Fig. \ref{fig8}
below) may be considered as pairs of virtually non-interacting stable
fundamental solitons supported by isolated nonlinear circles -- essentially
the same fundamental modes which were reported in Refs. \cite{HS} and \cite%
{Barcelona-OL}. On the other hand, the further increase of $a$ and/or
decrease of $R$ lead to the destabilization of the symmetric soliton --
first, through its conversion into the breather (see Fig. \ref{fig4}), which
is followed by the sudden transformation of the breather into a strongly
asymmetric single-peak soliton (see Fig. \ref{fig5}). These transitions can
be readily understood: the increase of $a$ and/or decrease of $R$ lead to
the enhancement of the interaction between the two\ circles, which causes
the destabilization of the double-peak soliton. As concerns the stability
border of the stable symmetric solitons, it has been checked that it is
precisely determined by the VK criterion, i.e., it coincides with the locus
of points where $dN/d\mu =0$ [recall, however, that these points do not
always determine a stability border, as soliton families which include
portions with $dN/d\mu <0$ may be unstable, see Fig. \ref{fig2}(a)].

\subsubsection{Antisymmetric solitons}

\begin{figure}[tbp]
\centering\subfigure[]{\includegraphics[width=3.5in]{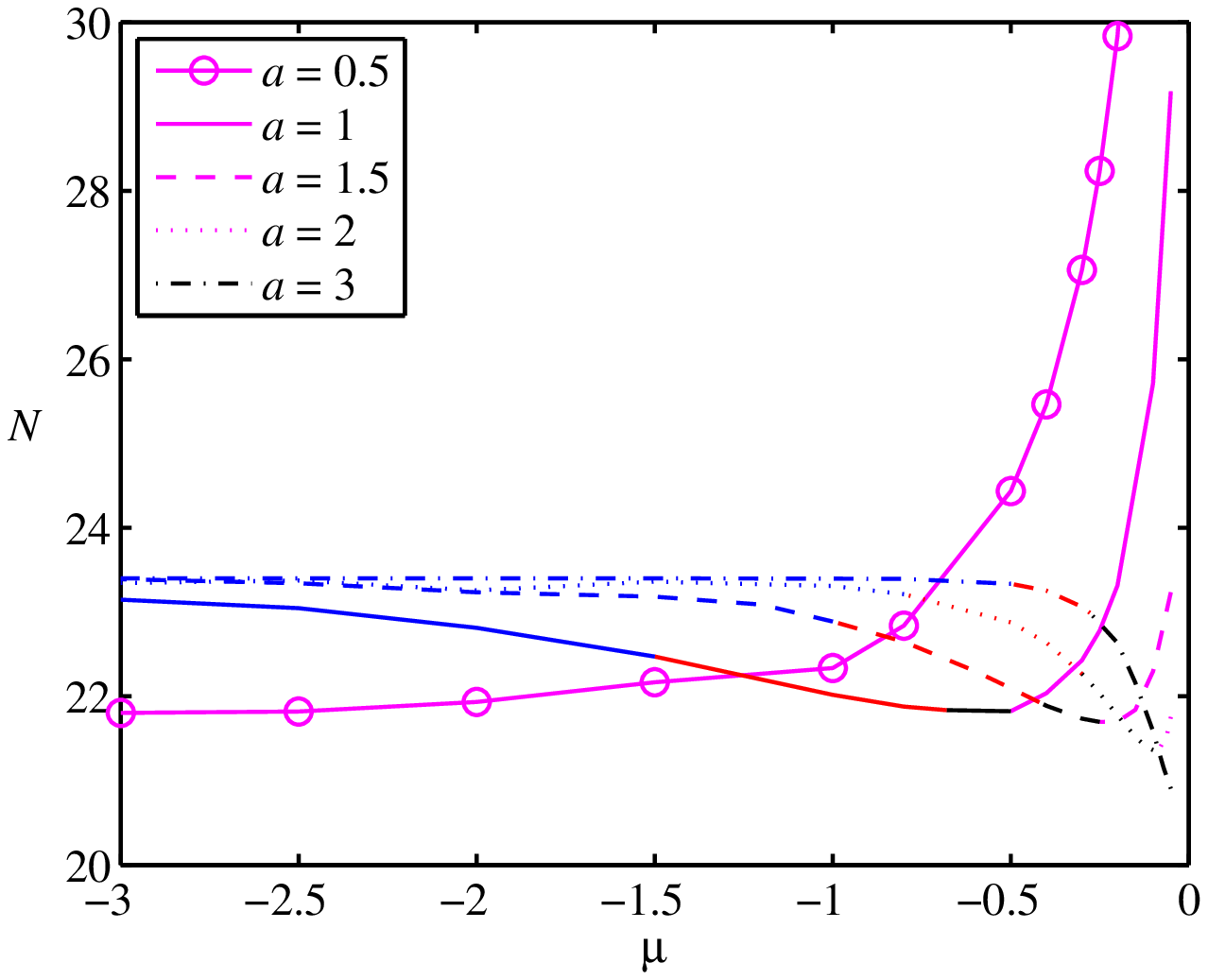}}%
\subfigure[]{\includegraphics[width=3.5in]{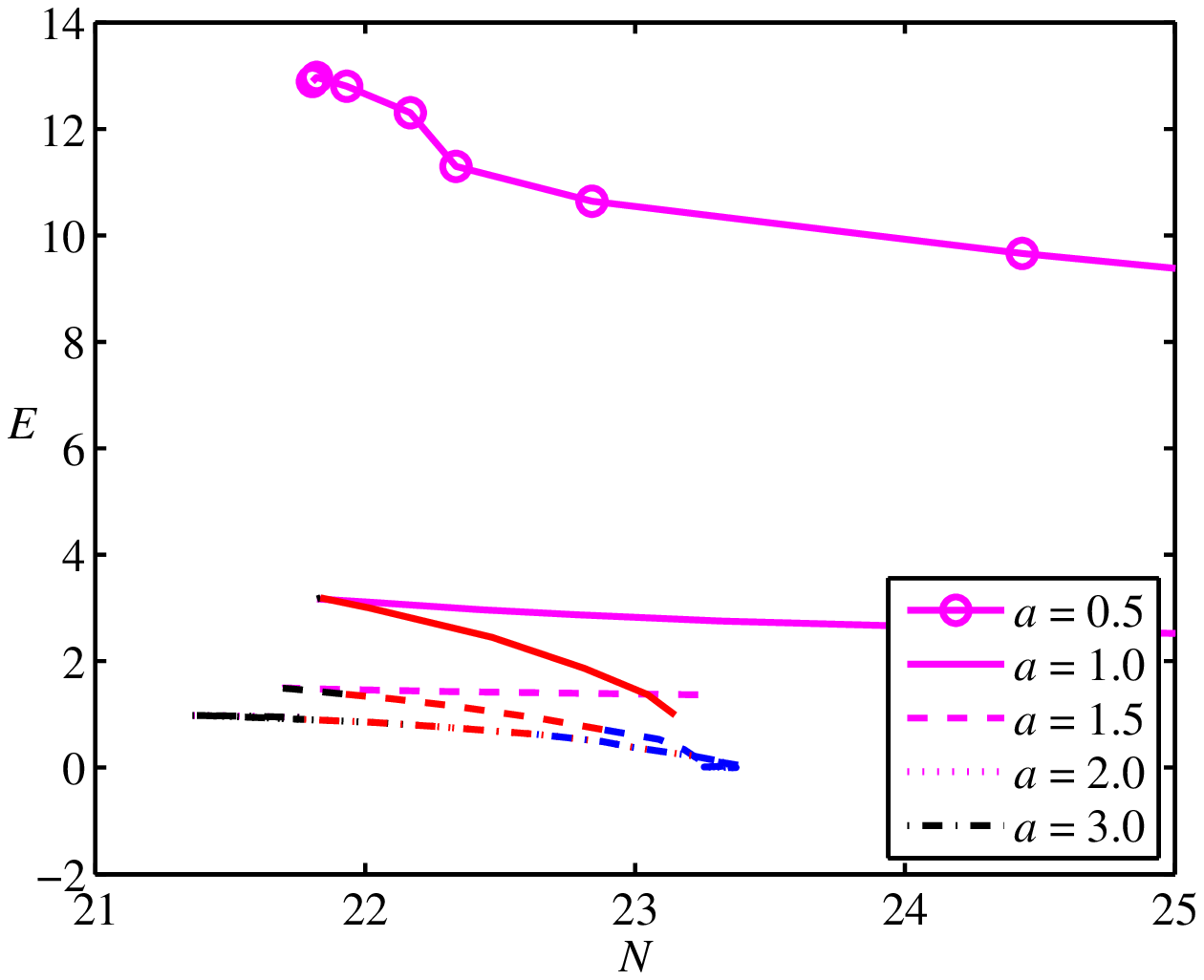}}
\caption{(Color online) The same as in Fig. \protect\ref{fig2} (for the
fixed separation between the circles, $L=3\protect\pi $), but for families
of antisymmetric stationary solitons.}
\label{fig7}
\end{figure}

\begin{figure}[tbph]
\centering\includegraphics[width=4in]{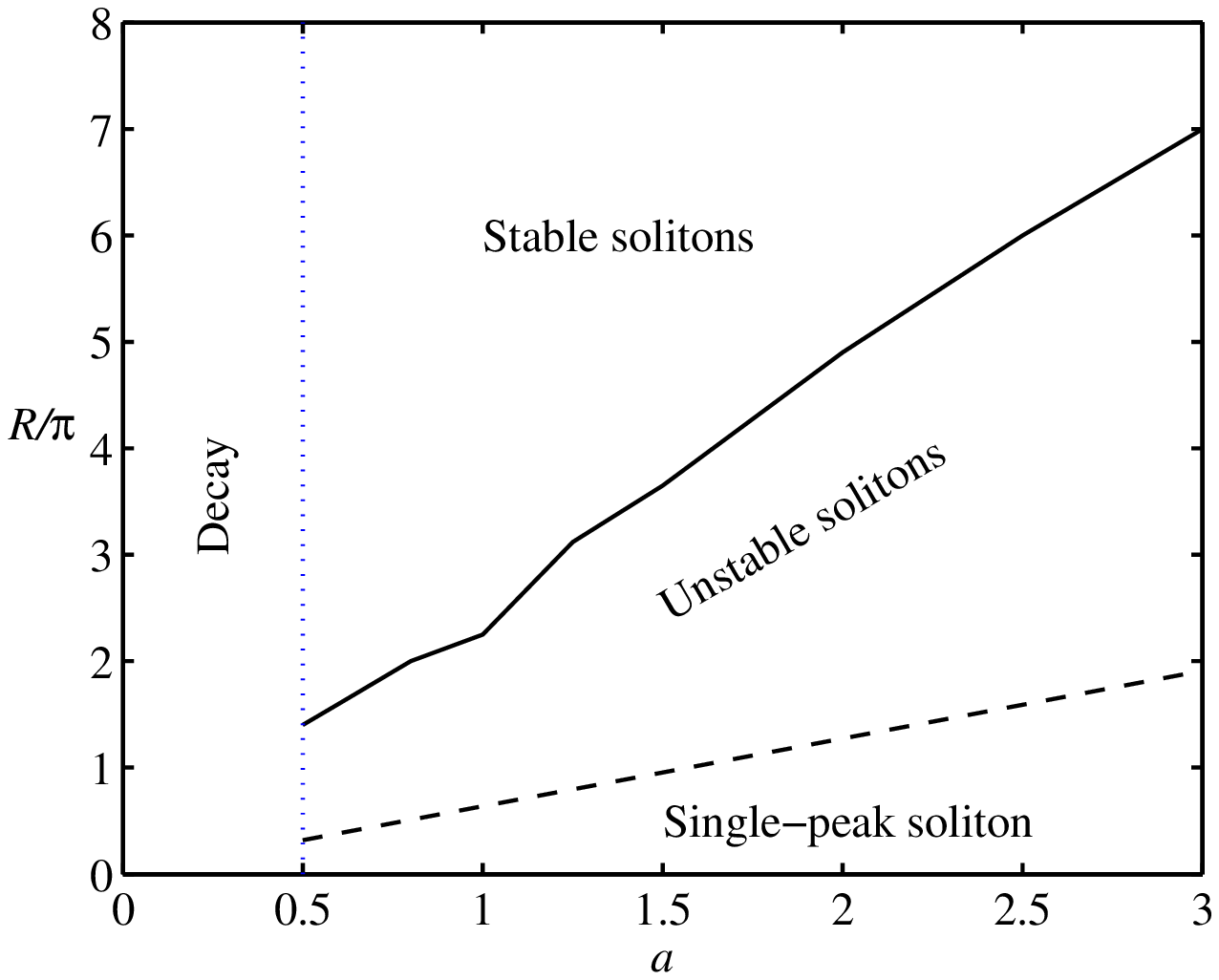}
\caption{(Color online) The same as in Fig. \protect\ref{fig6} (with the
fixed norm $N=22$), but for antisymmetric solitons and modes generated by
the development of their instability.}
\label{fig8}
\end{figure}

Figures \ref{fig7} and \ref{fig8} are counterparts of the above figures \ref%
{fig2} and \ref{fig6} for antisymmetric solitons, with $\phi \left(
x,y\right) =-\phi \left( -x,y\right) $. Predictions of the VA for
antisymmetric solitons are not included into Fig. \ref{fig7}(a), as, on the
contrary to the symmetric solitons, the agreement between the VA and the
numerical findings is poor in this case. The diagram in Fig. \ref{fig7}
includes a region of stable antisymmetric solitons, as well as an area
(``unstable solitons") where the unstable antisymmetric soliton
spontaneously transforms itself into a single-peak mode trapped in either
circle, and shedding off about a half of the initial norm. In terms of the
evolution of the respective density profile, $\left\vert \psi \left(
x,y,t\right) \right\vert ^{2}$, the latter outcome of the evolution is very
similar to that shown in Fig. \ref{fig5}.

Note that only parts of the solution branches with the negative slope, $%
dN/d\mu <0$, which meet the VK criterion, are stable in Fig. \ref{fig7}(a)
(see the curves pertaining to $a=1,1.5,~2$, and $3$). The stability and
instability intervals for these three values of $a$ are reported in Table 1.
Further, similar to the situation for the symmetric solitons displayed in
Fig. \ref{fig2}(b), the $E(N)$ curves for the antisymmetric solutions
displayed in Fig. \ref{fig7}(b) confirm the natural expectation that stable
portions of the solution families have lower energy than their unstable
counterparts, for $a=1$ and $1.5$.

\begin{table}[th]
\caption{Stability intervals, in terms of the total norm ($N$) and chemical
potential ($\protect\mu $), for families of stationary antisymmetric
solitons presented in Fig. \protect\ref{fig7}, for fixed $L=3\protect\pi $. }
\label{table:stability}\centering
\addtolength{\tabcolsep}{5pt}
\begin{tabular}{ccc}
\hline\hline
$a$ & $N$ & $\mu$ \\[0.5ex] \hline
1.0 & $21.8347 < N < 22.4702$ & $-1.5 <\mu< -0.68$ \\
1.5 & $21.9289 < N < 23.0877$ & $-1.2 <\mu< -0.42$ \\
2.0 & $22.1718 < N < 23.2130$ & $-0.8 <\mu< -0.3$ \\
3.0 & $22.9914 < N < 23.3356$ & $-0.5 <\mu< -0.278$ \\[1ex] \hline
\end{tabular}%
\end{table}

\subsection{Overlapping and touching circles ($L\leq 0$)}

As shown in Figs. \ref{fig6} and \ref{fig8}, stable double-peak solitons,
symmetric or antisymmetric ones, cannot be supported by the pair of circles
with a small separation between them. Further numerical analysis
demonstrates that such solitons cannot be found \ either for touching ($L=0$%
) or overlapping ($L<0$) circles. Instead, in this case it is possible to
find symmetric \emph{single-peak} solitons, centered at the midpoint of the
configuration. Strongly asymmetric solitons, trapped near the center of
either circle, while the other one is left almost empty, exist in this case
too, and they may be stable. These modes are considered below.

\subsubsection{Touching circles ($L=0$)}

\begin{figure}[tbp]
\centering\subfigure[]{\includegraphics[width=3.5in]{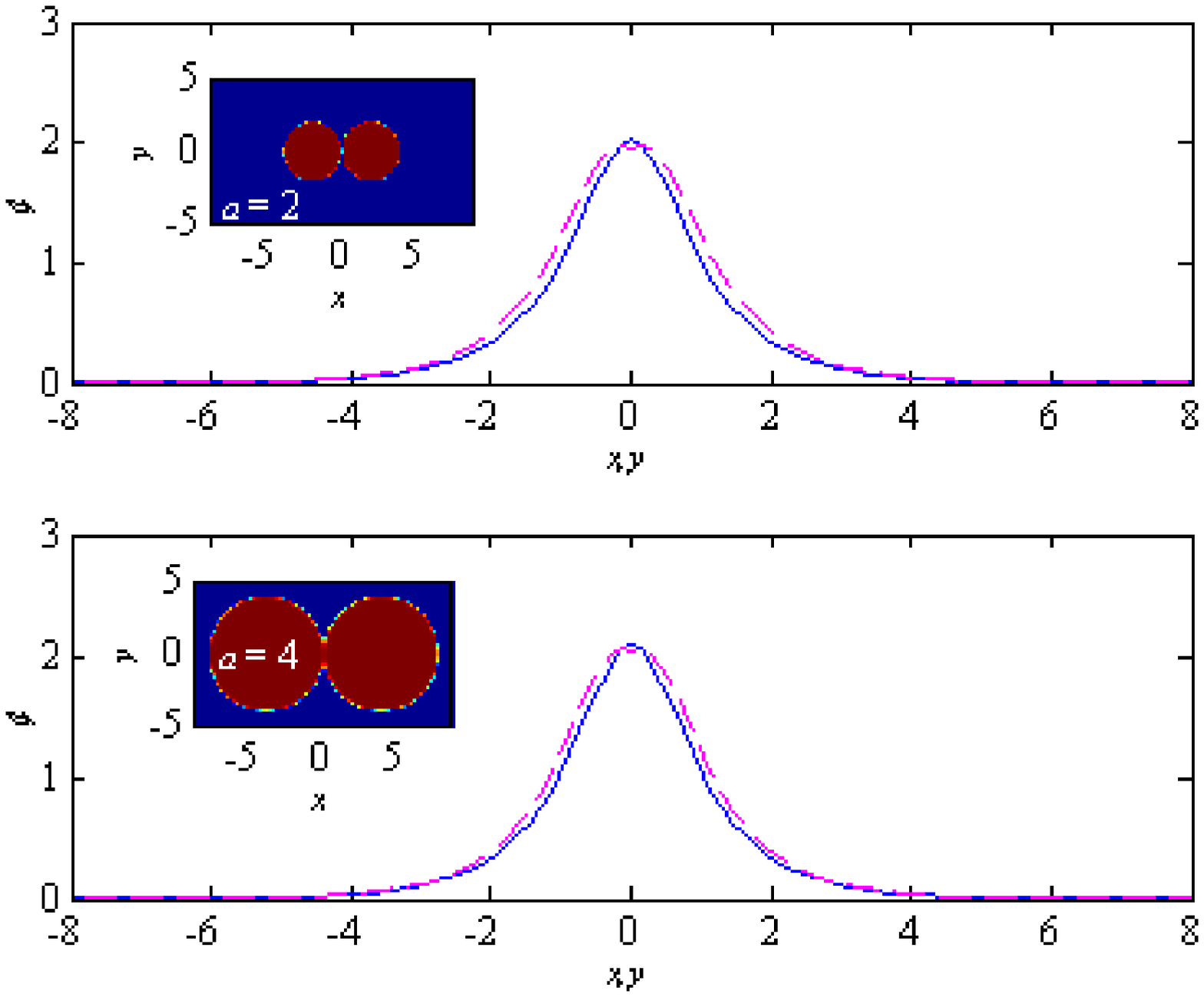}}%
\subfigure[]{\includegraphics[width=3.5in]{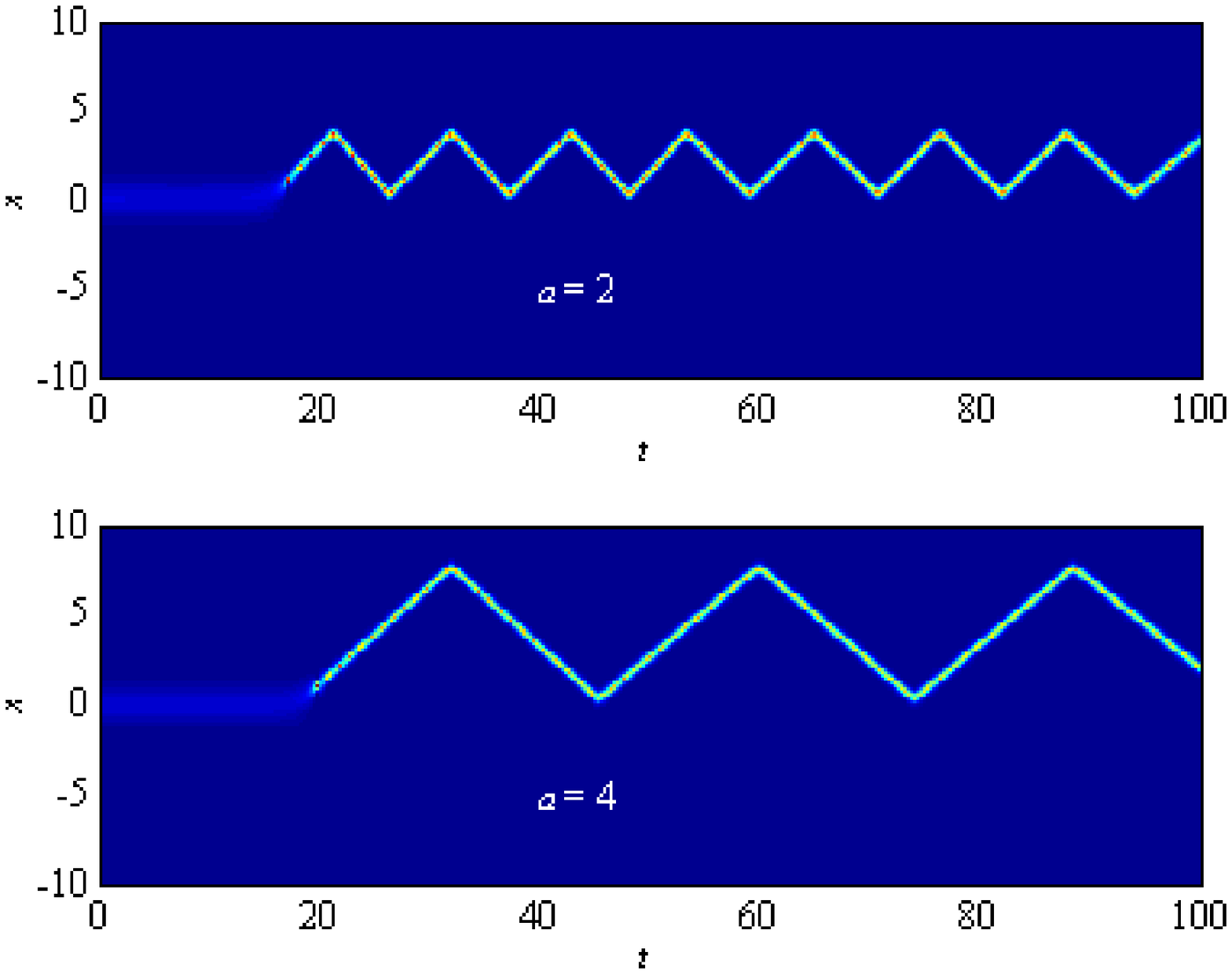}}
\caption{(Color online) (a) Profiles of unstable symmetric single-peak
solitons centered around the touch point of the set of two circles with zero
separation, $L=0$ (see the insets). The top and bottom panels pertain to
different radii of the circles, $a=2$ and $a=4$. In both cases, the norm of
the soliton is $N=15.9$. (b) The ensuing hopping motion of the single-peak
breather inside one circle, shown \ by means of the density contour plots in
the $x$-cross section.}
\label{fig9}
\end{figure}

In the case of $L=0$, the two circles touch each other at the single point,
featuring a ``figure of eight" [see insets in Fig. \ref{fig9}(a)]. As said
above, in this case symmetric stationary modes are represented by
single-peak solitons centered at the touch point, see examples of the
soliton profiles in Fig. \ref{fig9}(a). These solitons are unstable,
spontaneously jumping into one of the circles, as seen in Fig. \ref{fig9}%
(b). This instability can be easily understood, as the symmetric soliton is
actually located at the position of an unstable equilibrium between two
attractive pseudopotential wells corresponding to the circles. After the
jump, the single-peak soliton keeps hopping in the circle, periodically
hitting its border and bouncing back.

\begin{figure}[tbp]
\centering\includegraphics[width=2.5in]{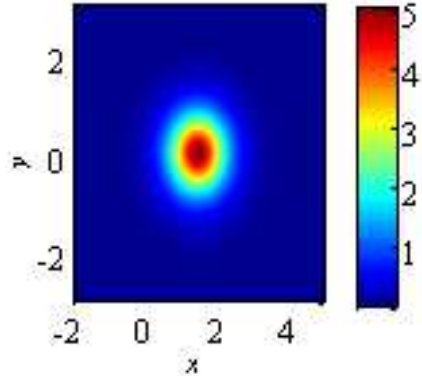}
\caption{(Color online) An example of the \emph{stable} single-peak soliton,
trapped \ in the right circle of the ``figure-of-eight" structure (with $L=0$%
), is shown by means of contour plots of the density, $\left( \protect\phi %
\left( x,y\right) \right) ^{2}$, for $\protect\mu =-1$, $N=11.4798$, and $%
a=1.5$.}
\label{fig10}
\end{figure}

Strongly asymmetric single-peak solitons trapped near the center of either
circle can be found too at $L=0$. An example of such a \emph{stable} nearly
isotropic soliton in shown in Fig. \ref{fig10}. The numerical data
demonstrate that the ratio of local densities at the center of the circle,
and at the midpoint where the two circles touch each other, is $\left[ \phi
\left( 1.5,0\right) /\phi \left( 0,0\right) \right] ^{2}\simeq 25$ for this
soliton.

\begin{figure}[tbph]
\centering\subfigure[]{\includegraphics[width=3.5in]{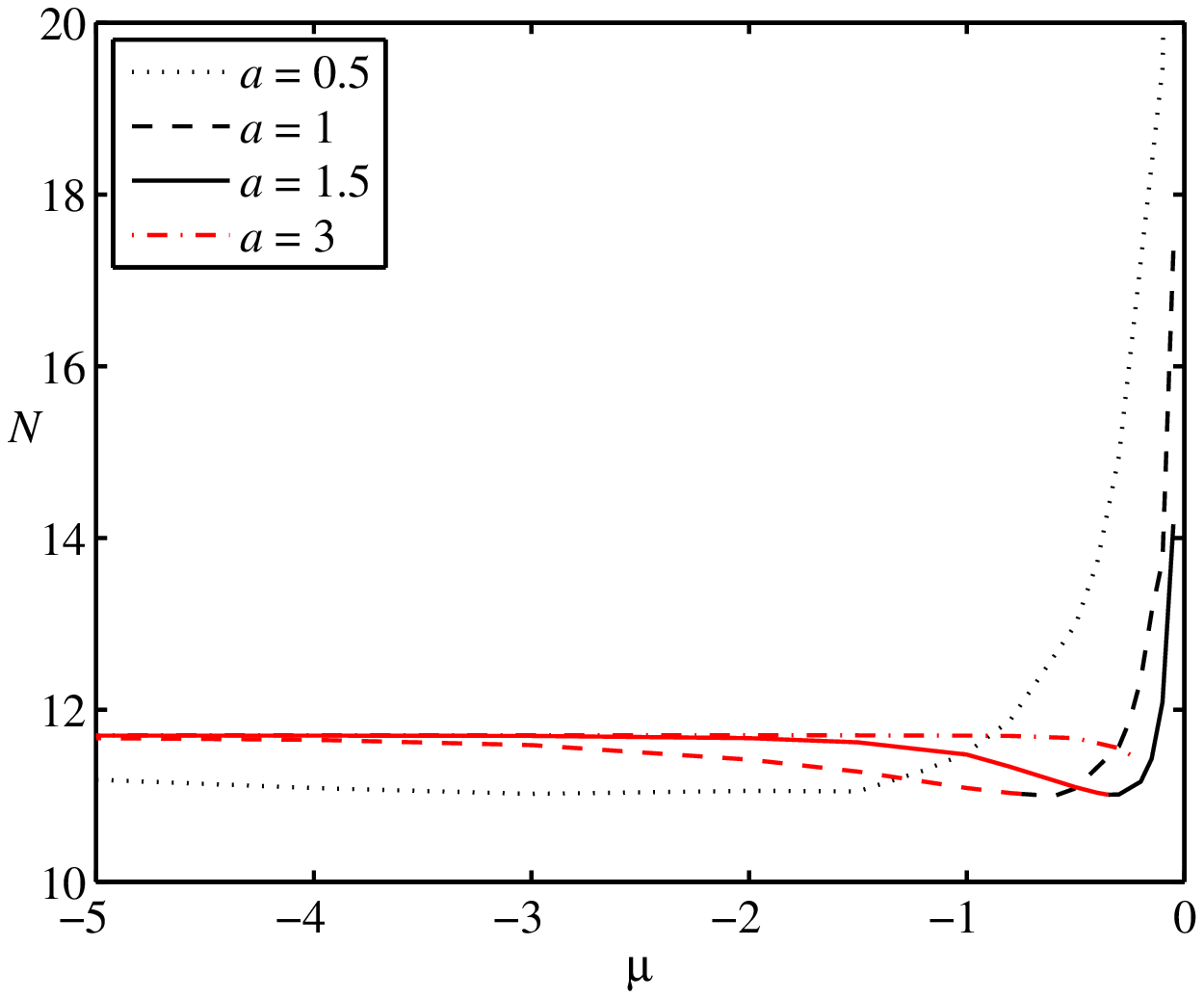}}%
\subfigure[]{\includegraphics[width=3.5in]{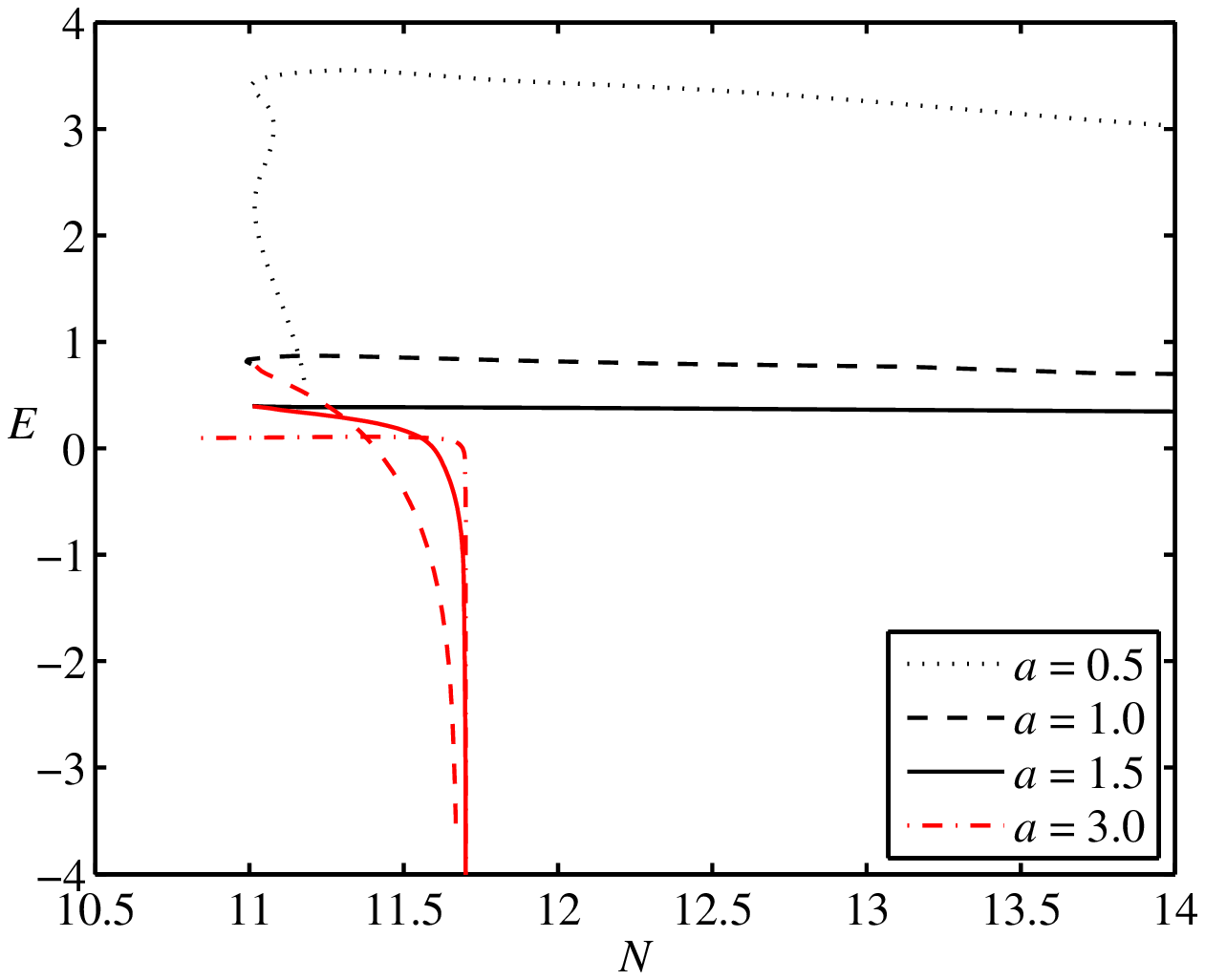}}
\caption{(Color online) The same as in Figs. \protect\ref{fig2}, but for the
family of stationary single-peak solitons nested in either circle of the
``figure-of-eight" configuration (the one with $L=0$). As before, the red
color designates stable solutions, while black branches represent modes
which are unstable through the decay into radiation.}
\label{fig11}
\end{figure}

Figure \ref{fig11} displays $N(\mu )$ and $E(N)$ dependences for families of
such single-peak asymmetric modes, obtained at different values of radius $a$%
. It is seen that stable subfamilies are found for $1\leq a\leq 3$. The
stability of these solution families \emph{exactly} obeys the VK criterion,
with unstable modes featuring decay into radiation (not shown here).
Extremely close to the critical value (\ref{Townes}), the stationary
single-peak solitons develop a weak oscillatory instability and turn into
breathers featuring intrinsic regular vibrations with a small amplitude (not
shown here in detail). For instance, at $a=1.5$, the robust breathers are
observed in an interval of $11.668<N<11.690$.

\subsection{Partly overlapping circles ($L<0$)}

\begin{figure}[tbp]
\centering\subfigure[]{\includegraphics[width=3.5in]{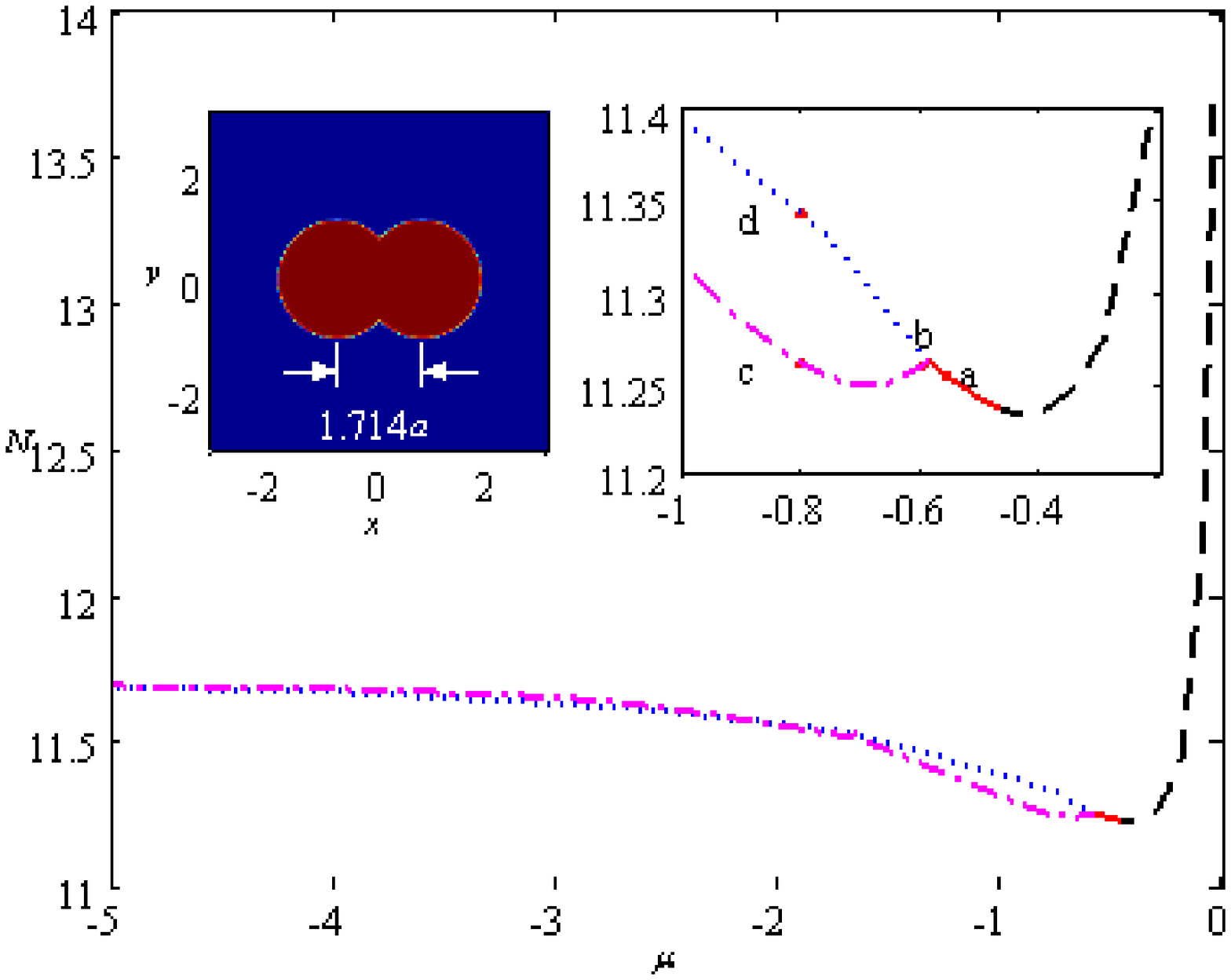}}%
\subfigure[]{\includegraphics[width=3.5in]{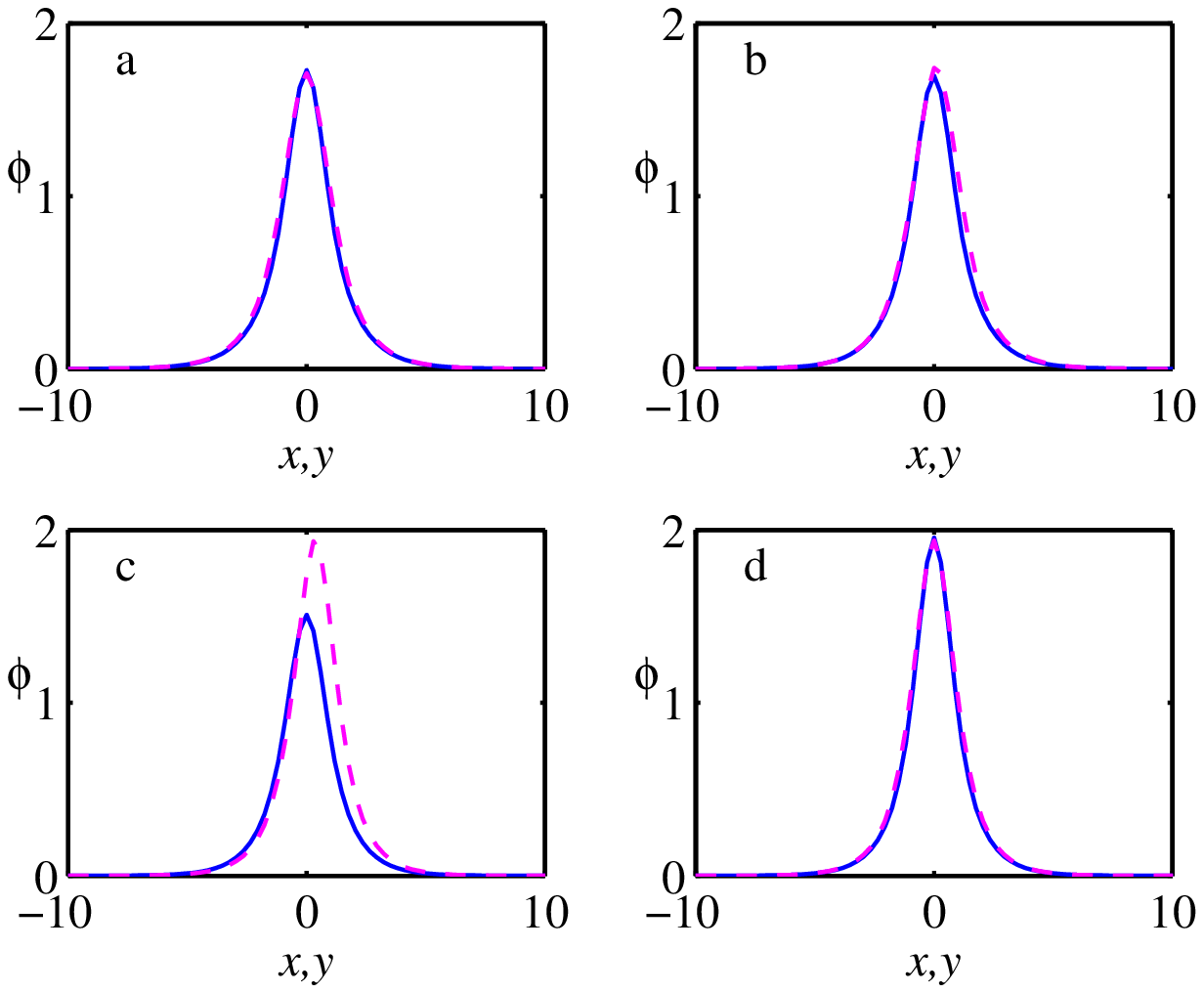}}
\caption{(Color online) (a) The $N(\protect\mu )$ curves for stationary
single-peak solitons in the configuration with $L=-0.5720a$ and $a=1.1$. In
part (b), subplots labeled ``a", ``b", ``c" and ``d" display cross sections
along $x $ (the magenta dashed lines) and along $y$ (the blue solid lines)
of the single-peak solitons corresponding to points ``a", ``b", ``c" and
``d" in panel (a), i.e., respectively, for $\protect\mu =-0.56$, $N=11.2544$%
; $\protect\mu =-0.6$, $N=11.2592$; $\protect\mu =-0.8$, $N=11.2606$; and $%
\protect\mu =-0.8 $, $N=11.3421$.}
\label{fig13}
\end{figure}

Negative $L$\ ($-2a<L<0$) implies that the two circles partly overlap [see
the top left inset in Fig. \ref{fig13}(a)], merging into a single circle at $%
L$\ $=$ $-2a$. Only single-peak solitons were found in this case. First, we
continue the numerical analysis of the symmetric solitons centered at the
midpoint, $x=y=0$. In a narrow interval corresponding to slightly
overlapping circles, $-0.1a\leq L<0$, the symmetric solitons are unstable in
essentially the same fashion as demonstrated above in Fig. \ref{fig9}, i.e.,
they jump from the unstable equilibrium position into either circle,
performing the shuttle motion in it.

As the degree of the overlap between the circles increases, in the interval
of $-0.33a\leq L<-0.1a$ the unstable symmetric soliton simply decays into
radiation (not shown here in detail). This outcome of the evolution may be
interpreted as a consequence of the fact that the total norm of the soliton
is too low in this case, not allowing it to maintain its integrity. Further,
keeping to increase the overlap degree, in the interval of $-0.33a<L\leq
0.48 $ we observe that the gradual enhancement of the nonlinearity around
the midpoint does not yet stabilize the symmetric soliton, but restores its
integrity. In this case, the soliton again spontaneously leaps to the left
or right circle, which is followed by its shuttle motion inside the circle.

The symmetric solitons centered around the midpoint may be stable at $L\leq
-0.48a$. For instance, at $L=-0.50a$, they are stable in the interval of $%
N_{\min }=11.2523<N<11.2665=N_{\max }$, which corresponds to $-0.55<\mu
<-0.47$, cf. Table 1. Similarly, at $L=-0.55a$\ the stability interval is $%
11.2375<N<11.2637$, or $-0.59<\mu <-0.47$. The stability interval for the
symmetric solitons expands with the increase of $|L|$ towards $L=-2a$.

The situation when the symmetric solitons may be stable is illustrated by
Fig. \ref{fig13}. The right inset to panel \ref{fig13}(a) demonstrates that,
at $\mu =-0.6$\ ($N=11.2592$), the $N(\mu )$\ curve splits into two
branches, which then merge at $\mu =-1.65$\ ($N=11.5295$). The resulting
bifurcation loop resembles those found in the 1D and 2D double-core models
with the cubic-quintic nonlinearity \cite{CQ,CQ3}. The splitting means that
two soliton solutions with different values of the norm can be found at
given $\mu $ (for instance, the pair of solitons corresponding to points
``c" and ``d" in Fig. \ref{fig13}). In the splitting region, the top branch
in panel \ref{fig13}(a) represents the symmetric soliton placed at the
midpoint, while the lower branch represents a soliton shifted to left or
right. In fact, the splitting and recombination of the two solution branches
are explicit examples of the direct and inverse symmetry-breaking
bifurcations.

Note that the solitons of types ``a" and ``d", shown in Fig. \ref{fig13}(b),
are centered exactly at the midpoint, while the solitons of types ``b" and
``c" show a small shift off the midpoint. Of course, together with the
shifted soliton, it is possible to find its mirror-image counterpart,
shifted in the opposite direction.

The simulations demonstrate that broad symmetric solitons, corresponding to
large values of $N$ [cf. Eq. (\ref{asympt})], suffer decay at $\mu >-0.47$\
(i.e., at $N>11.2361$), as indicated by the dashed black line in Fig. \ref%
{fig13}(a). Stable symmetric solitons where found in the interval of $%
-0.6<\mu <-0.47$\ (in terms of the norm, it is $11.2654<N<11.2361$), which
corresponds to the short solid red segment in Fig. \ref{fig13}(a). At $\mu
<-0.6$, the solitons marked by ``b" and ``c", which belong to the family
indicated by the dashed-dotted magenta curve, spontaneously transform
themselves into breathers which feature regular small-amplitude
oscillations. Further, solitons pertaining to the dotted blue curve in Fig. %
\ref{fig13}(a) (for instance, the soliton corresponding to point ``d" in
Fig. \ref{fig13}) undergo a spontaneous transformation into breathers which
feature strong irregular intrinsic oscillations, but remain robust localized
modes.

\begin{figure}[tbph]
\centering\includegraphics[width=4in]{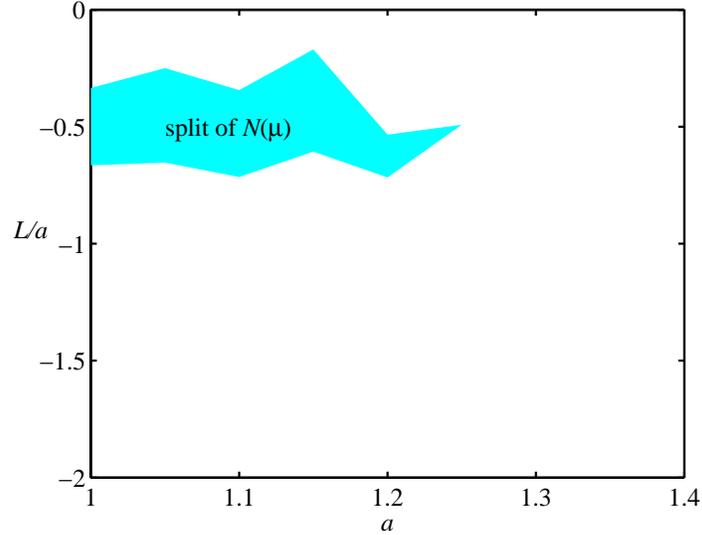}
\caption{(Color online) The region in the plane of $\left( a,L/a\right) $,
at fixed $\protect\mu =-0.8$, where curves $N(\protect\mu )$ for the family
of single-peak solitons, supported by the overlapping circles, feature the
splitting [see Fig. \protect\ref{fig13}(a)].}
\label{fig17}
\end{figure}

The splitting of the $N(\mu )$\ curve occurs in a small region of the plane
of $\left( a,L/a\right) $, as shown in Fig. \ref{fig17}. As concerns the
(narrow) stability area of the symmetric solitons, it is outlined by Table
2, for the overlapping parameter fixed with respect to the radius of the
circles, $L=-0.5720a$ (the same relation between $L$ and $a$ as in Fig. \ref%
{fig13}). At values of $N$ exceeding the largest value indicated in Table 2,
the symmetric solitons spontaneously turn into robust single-peak breathers.

\begin{table}[th]
\caption{Stability intervals for families of stationary symmetric solitons
centered around the midpoint of the configuration with partly overlapping
circles, $L=-0.5720a$, and several values of radius $a$ of the overlapping
circles. }
\label{table:stability2}\centering
\addtolength{\tabcolsep}{5pt}
\begin{tabular}{ccc}
\hline\hline
$a$ & $N$ & $\mu$ \\[0.5ex] \hline
1.1 & $11.2361 < N < 11.2620 $ & $-0.47 < \mu < -0.588$ \\
1.5 & $11.3449 < N < 11.2633 $ & $-0.33 < \mu < -0.47$ \\
1.8 & $11.2025 < N < 11.2611 $ & $-0.17 < \mu < -0.25$ \\
$a\geq 2$ & no stability intervals &  \\[1ex] \hline
\end{tabular}%
\end{table}

Finally, for strongly overlapping circles, with $-2a\leqslant L<-0.8$, the
results, including the stability region for the symmetric solitons (only the
symmetric ones exist in this case) are nearly the same as reported in Ref.
\cite{HS} for the single circle, which corresponds to $L=-2a$.

\section{Summary}

In this work, we have introduced the 2D model based on the modulation
function confining the self-focusing cubic nonlinearity, i.e., the
corresponding nonlinear potential, to the two identical circles, which may
be separated or overlapped. We aimed to study symmetric, antisymmetric, and
asymmetric stationary 2D solitons in this setting, as well as breathers into
which unstable solitons may be spontaneously transformed. Results of the
analysis, obtained by means of numerical methods and the VA (variational
approximation), can be summarized as follows below.

Well-separated circles support stable symmetric and antisymmetric solitons,
as long as the interaction between the density peaks trapped in the two
circles is negligible. The increase of the interaction strength (i.e., the
decrease of separation $L$ between the circles, relative to their radii $a$)
leads to the destabilization of the symmetric and antisymmetric solitons. As
shown in Fig. \ref{fig6}, the symmetric solitons pass two SSB transitions.
At first, they are transformed into robust breathers featuring
small-amplitude irregular intrinsic vibrations, with a spontaneously
emerging small asymmetry between the two peaks (Fig. \ref{fig4}). As the
interaction between the two circles grows still stronger, the weakly
asymmetric double-peak breathers lose their stability and are replaced by
single-peak modes residing in one circle, while the other one remains nearly
empty. On the other hand, Fig. \ref{fig8} shows that the antisymmetric
solitons feature the single SSB transition, by a direct jump into the
single-peak mode. The VA, based on the superposition of two Gaussians
centered in the two circles, describes the symmetric solitons reasonably
well, while it fails to accurately approximate antisymmetric ones.

The situation is different in the case of touching ($L=0$) and overlapping ($%
L<0$) circles. In this case, only single-peak solitons are found -- both
asymmetric modes, trapped in one circle (Fig. \ref{fig10}), a part of which
are stable (Fig. \ref{fig11}), and symmetric solitons centered around the
midpoint ($x=y=0$). For the weak overlapping, $0\leq -L\leq 0.1a$, the
symmetric single-peak soliton is situated at the unstable-equilibrium
position. As a result, it spontaneously leaps into either circle and then
performs shuttle motion therein (Fig. \ref{fig9}). For a stronger overlap, $%
0.1a<-L<0.33a$, the symmetric soliton decays into radiation. At a still
larger degree of the overlap, $0.48a<-L<0.33a$, the enhancement of the local
nonlinearity around the midpoint results in the reappearance of the SSB
regime, in the form of the spontaneous leap followed by the shuttle motion.
A region of the stability of the symmetric single-peak solitons appears at $%
-L>0.48a$, gradually expanding, with the growth of $|L|$, into the stability
region for the soliton trapped in the single nonlinear circle \cite{HS} in
the limit when the two circles merge into the single one ($-L\rightarrow 2a$%
). In the case of the moderately strong overlap, a noteworthy manifestation
of the SSB was found in the form of the paired symmetry-breaking and
symmetry-restoring bifurcations, which give rise to the loop in Fig. \ref%
{fig13}(a).

This work may be naturally extended in other directions. In particular, a
related dynamical problem is \textit{Josephson oscillations} in bosonic
junctions formed by double-well potentials, cf. Refs. \cite{mean-field},
\cite{Salerno}-\cite{Mazzarella}. The present model calls for the study of
Josephson oscillations in the double-well \textit{nonlinear} \textit{%
pseudopotentials}. Because the setting has no linear limit, the Josephson
frequency is expected to strongly depend on the norm of the condensate. The
models considered in Ref. \cite{we} and here suggest that the corresponding
\textit{nonlinear Josephson junctions} may be experimentally realized in
photonic media, in addition to BEC.

The nonlinear pseudopotentials may be also naturally generalized for
two-component models, which give rise to vector solitons \cite%
{Barcelona-vector}. Accordingly, it may be relevant to study SSB effects in
double-well pseudopotentials trapping two-component mixtures.

\section*{Acknowledgment}

We appreciate valuable discussions with M. Salerno. The work of T.M. was
supported by the Thailand Research Fund under grant RMU5380005.

\end{document}